\documentclass[twocolumn,showpacs,showkeys,preprintnumbers,
               amsmath,amssymb,floatfix,10pt]{revtex4}
\usepackage{graphicx}% Include figure files
\usepackage{dcolumn}% Align table columns on decimal point
\usepackage{bm}% bold math
\usepackage{graphicx}
\usepackage{subfigure}
\usepackage{amsmath}
\usepackage{amssymb}
\usepackage{wrapfig}
\usepackage{color}
\usepackage[english]{babel}
\usepackage[warning,math]{easyeqn}
\usepackage[thinlines]{easybmat}
\newcommand{\ds}{\displaystyle}
\newcommand{\scs}{\scriptscriptstyle}

\newcommand{\de}{\partial}
\def\g{\gamma}

\def\o{\omega}

\def\<{\langle}
\def\>{\rangle}
\begin{document}
%
%%%%%%%%%%%%%%%%%%%%%%%%%%%%%%%%%%%%%%%%%%%%%%%%%%%%%%%%%%%%%%%%%%%%%%%%%%%
\preprint{}
%
%%%%%%%%%%%%%%%%%%%%%%%%%%%%%%%%%%%%%%%%%%%%%%%%%%%%%%%%%%%%%%%%%%%%%%%%%%%
\title{Slow energy relaxation of macromolecules and nano-clusters in solution} 
\author{F. Piazza} 
\author{P. De Los Rios}
\affiliation{Institute of Theoretical Physics-LBS,
Ecole Polytechnique F\'ed\'erale de Lausanne (EPFL),  
CH-1015 Lausanne, Switzerland\email{Francesco.Piazza@epfl.ch}}  
%
%%%%%%%%%%%%%%%%%%%%%%%%%%%%%%%%%%%%%%%%%%%%%%%%%%%%%%%%%%%%%%%%%%%%%%%%%%%%
%
\author{Y.--H. Sanejouand}
\affiliation{Laboratoire de Physique,
Ecole Normale Sup\'erieure, 46 all\'ees d Italie,   
69364 Lyon Cedex 07, France}
%
%---------------------------------------------------------------------------
\begin{abstract} 

Many systems in the realm of nanophysics from both the living and inorganic
world display slow relaxation kinetics of energy fluctuations. 
In this paper we propose a general  explanation for such phenomenon,
based on the effects of interactions with the solvent.
Within  a simple harmonic model of the system fluctuations, we demonstrate 
that the inhomogeneity of coupling to the solvent of the bulk and 
surface atoms suffices to generate a complex spectrum of decay rates.
We show for Myoglobin and for a metal nano-cluster that 
the result is a complex, non-exponential relaxation dynamics.

\end{abstract} 
%
%%%%%%%%%%%%%%%%%%%%%%%%%%%%%%%%%%%%%%%%%%%%%%%%%%%%%%%%%%%%%%%%%%%%%%%%%%%%
%---------------------------------------------------------------------------
%
\pacs{05.70.Ln; 87.15.-v; 61.46.+w}
%
%------------------------------- SOME RELATED PACS ---------------------------
%
%  05.70.Ln 	Nonequilibrium and irreversible thermodynamics	
%  82.60.Qr 	Thermodynamics of nanoparticles
%  61.46.+w 	Nanoscale materials: clusters, nanoparticles, nanotubes, 
%               and nanocrystals 
%  65.80.+n 	Thermal properties of small particles, nanocrystals, nanotubes
%  87.15.-v 	Biomolecules: structure and physical properties
%  87.15.He 	Dynamics and conformational changes
%  87.15.Ya 	Fluctuations
%
%--------------------------------------------------------------------------------
%
\keywords{Slow relaxation, proteins, nano-clusters}
\maketitle
%
%%%%%%%%%%%%%%%%%%%%%%%%%%%%%%%%%%%%%%%%%%%%%%%%%%%%%%%%%%%%%%%%%%%%%%%%%%%%%%%%%%
%

Matter from both the living and the inorganic worlds displays unusual structural 
and dynamical properties when  it is reduced to small objects of nano-metric scale ~\cite{nanoworld}. 
One interesting feature that has recently been arousing interest is the relaxation to thermal 
equilibrium of local or
distributed temperature fluctuations. 
Heat dissipation experiments in metallic nano-clusters in solution
have revealed  that the approach to equilibrium follows a stretched-exponential law~\cite{slow.clus}. Similarly, experiments on proteins  have shown that their relaxation after
excitation may also be  highly non--exponential~\cite{frauenfelder,slow.prot}. 

Slow energy relaxation is naturally associated with the presence 
of a hierarchy  of relaxation times. Nevertheless,  the origin of such hierarchy
is not always easy to pinpoint. A common assumption
is that multiple time scales arise because of the roughness
of the energy landscape, {\em i.e.} lots of local minima of different depths
whose sequence {\em en route} to equilibrium determines 
a cascade of decay times. However, while such a picture would be
compatible with our current understanding of protein folding dynamics~\cite{dill},
such mechanism is unlikely to apply to metallic nano-clusters, which are regular stacks 
of identical  atoms, and whose energy landscape close to the ground state is likely to be mostlyfeatureless. Yet, the relaxation dynamics in solution of proteins and nano-clusters
share enough similarities to raise the question whether a common mechanism could be at work 
in a broad class of nano-systems.

The physiological environment of proteins  is a viscous aqueous solution, 
while metallic nano-clusters are believed to be  promising markers to study the
behavior of cells and living tissues, so that an aqueous solution is also the environment in many of
their applications.   In this paper we propose that a natural, simple and unifying explanation for the
spectrum of decay times for nano-particles in solution can be found in the different coupling to the
solvent of surface and bulk atoms. 

%%%%%%%%%%%%%%%%%%%%%%%%%%%%%%%%%%%%%%%%%%%%%%%%%%%%%%%%%%%%%%%%%%%%%%%%%%%%%%%%%%
\begin{figure}[b!]
\centering
\includegraphics[width=9. truecm,clip]{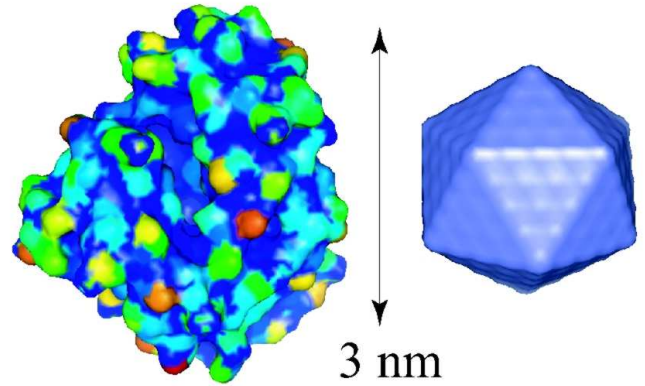}
\caption{\small \label{f:strutt} Structure of Myoglobin (pdb code 1A6M) and 
of a model icosahedral metal cluster with $N=309$ atoms. 
The protein structure is color--coded
according to the fraction of surface accessible to the solvent at each site --
from blue (low accessibility) to red (highest exposure).}
\end{figure}
%%%%%%%%%%%%%%%%%%%%%%%%%%%%%%%%%%%%%%%%%%%%%%%%%%%%%%%%%%%%%%%%%%%%%%%%%%%%%%%%%%

The idea is that energy is dissipated to the environment only 
by  surface atoms, whereas bulk atoms only exchange energy with each other with very little or 
no dissipation. To investigate the effects of the damping inhomogeneity, 
we model the systems under investigation as elastic networks. The interatomic interactions are simple 
harmonic potentials $V(\vec{r}_i,\vec{r}_j)=k_{ij}/2 (|\vec{r}_i-\vec{r}_j|-
|\vec{r}_{i0}-\vec{r}_{j0}|)^2$, where $\vec{r}_i$  is the position of atom $i$,
$\vec{r}_{i0}$ its equilibrium position and $k_{ij}$ is the
interaction stiffness between atoms $i$ and $j$. 
In the case of metallic nano-clusters this modelization is the usual, textbook description, 
which correctly predicts most of the general features of crystal
vibrations~\cite{Ashcroft}.  The
application of elastic network models to proteins is  more recent~\cite{Tirion}, since it had been
customary to assume that proteins  are characterized by complex  energy landscapes. Yet, it has been
realized that most features of the large- and medium-scale dynamics of proteins close to their native
state can be successfully reproduced by simple 
harmonic interactions between amino-acids~\cite{Bahar,hinsen,yhs}. 
In these
coarse-grained models $\vec{r}_i$ represents the position of the  $\alpha$-carbon of the $i$-th
amino-acid, $\vec{r}_{i0}$ its position in the native state as determined from X-ray crystallography or
Nuclear Magnetic Resonance, and $k_{ij}$ can take different functional forms, such as $k_{ij}=
k\; \theta(|\vec{r}_{i0}-\vec{r}_{j0}|-r_c)$ or $k_{ij}=k\; \exp(-|\vec{r}_{i0}-\vec{r}_{j0}|^2/r_c^2)$,
where $r_c$ is a suitable cutoff (or typical) interaction distance that tunes the overall connectivity
of the structure, and $k$ a phenomenological strength constant. 

Treating the systems as elastic networks is indeed the simplest hypothesis.
Yet, recent rigorous findings about energy relaxation in one and two dimensions
suggest that even linear distributed systems display highly non-trivial relaxation 
properties~\cite{Piazza1,Piazza2}. 
We thus start up illustrating the solution to the following problem:
a chain of $N$ equal masses connected by
springs of equal  strength, coupled to a heat bath at its edges. 
The chain has free ends and at $t=0$ is at equilibrium at a
temperature $T_{\scs 0}$. Modeling the system \`a la Langevin, only the first and last particles
are damped and subject to randomly fluctuating forces due to their interaction with the solvent,
according to the equations
\begin{eqnarray}
m \ddot{x_i} &=& k (x_{i-1} -2 x_i + x_{i+1}) \;\;\;\; i=2,...,N-1 \nonumber \\
m \ddot{x_1} &=& k (x_2 - x_1) - \gamma \dot{x_1} + \eta_1 \nonumber \\
m \ddot{x_N} &=& k (x_{N-1} - x_N) - \gamma \dot{x_N} + \eta_N 
\label{Langevin 1d}
\end{eqnarray}
where $\gamma$ is the viscous friction coefficient and $\eta_k(t)$ ($k=1,N$)
is a Gaussian, delta-correlated white noise, whose standard deviation
is fixed by the fluctuation dissipation theorem.
We want to study the system relaxation to a temperature $T<T_{\scs 0}$.

The time behavior of the system only depends on the temperature difference 
$\Delta T = T_{\scs 0} - T$ because of the absence of local energy barriers
in the energy landscape.
This property shall be proved rigorously later on in the framework 
of the Fokker--Planck (FP) approach.
As a consequence, we may solve the problem as that of a 
deterministic dissipative
system of energy $N k_{\scs B} \Delta T$ damped at its edges.
Such problem is amenable to a perturbative treatment in the limits of small and 
large damping,
resulting in a decay spectrum of the linear modes given by~\cite{Piazza1}
\begin{equation}
\label{dampspec1D}
\g(\o) = \g_{\scs 0} \left[ 
               1 - \left( \frac{\o}{\o_0} \right)^2
              \right] 
\,,	\ \     
\g_{\scs 0} =
\begin{cases}  
f\g        & \g \ll \o_0 \\
\frac{\ds f\o^2_0}{\ds 4\g}  & \g \gg \o_0
\end{cases}  
\end{equation}
where $\gamma$ and $\o_0$ are the damping strength and band--edge frequency,
respectively, and $f=2/N$ is the \emph{surface} fraction.
The superposition of decay constants of the form~(\ref{dampspec1D})
allows to calculate the energy relaxation analytically,
revealing a crossover in the system from the initial exponential
decay $\exp(-\g_{\scs 0}t)$ (single fastest relaxing mode) to a power--law relaxation of 
the type $t^{-1/2}$ ({\em integrated} regime)~\cite{Piazza1}.
Accordingly, the solution to the original problem may be written as 
\begin{equation}
\label{Bessel1D}
\begin{aligned}
\frac{\< E(t) \> - \< E(\infty) \>}{\< E(0) \> - \< E(\infty) \>} 
                       &= \int_0^\infty g(\o) e^{ -2\g(\o)t} \, d\o =  
                          \mathcal{I}_0 \left(
                                     \g_{\scs 0}t
                                     \right)  \\
                       &\approx
\begin{cases}
e^{\ds-\g_{\scs 0} t}           & t \ll 1/\g_{\scs 0} \\
(2\pi \g_{\scs 0}t)^{ -1/2}  & t \gg 1/\g_{\scs 0} 
\end{cases} 
\end{aligned}
\end{equation}
where $g(\omega)$ is the density of modes and 
$\mathcal{I}_0$ is the exponentially-modified zero--order Bessel function.
The same calculation may be repeated in higher dimensions, yielding 
\begin{equation}
\label{BesselD}
\frac{\< E(t) \> - \< E( \infty) \>}{\< E(0) \> - \< E(\infty) \>}
                           =  \left[ 
                               \mathcal{I}_0 \left(
                                     \g_{\scs 0}t
                                   \right)
                                \right]^d 
\end{equation}
where $d$ is the spatial dimension, the parameter $\g_{\scs 0}$ still 
carrying the information on the {\em linear} surface fraction $f/d$.
In the rest of the Letter we 
show that a similar  crossover from an exponential to an integrated, non--exponential 
regime also occurs for more complex geometries and 
connectivities. 
In order to treat the  general case, 
we numerically compute the exact solution of the FP formulation of the problem.

%%%%%%%%%%%%%%%%%%%%%%%%%%%%%%%%%%%%%%%%%%%%%%%%%%%%%%%%%%%%
\begin{figure*}[t!]
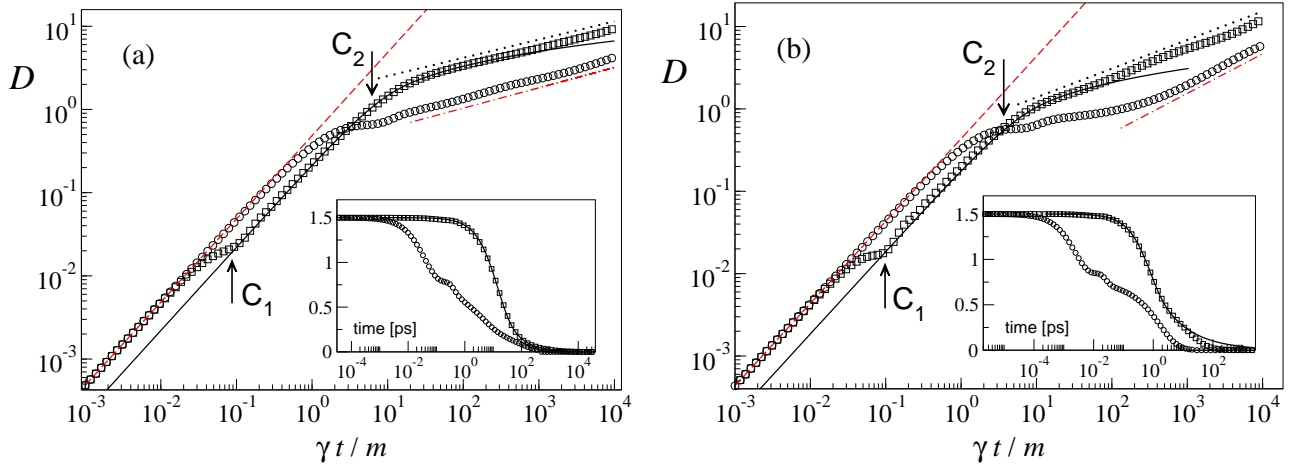

\centering
\subfigure{
\includegraphics[width=8.3 truecm,clip]{Fig2.eps}}
\hspace{0.5 mm}
\subfigure{
\includegraphics[width=8.3 truecm,clip]{Fig3.eps}}
\caption{\small Relaxation to equilibrium in Myoglobin (a) and 
in an icosahedral cluster (b).
Symbols are plots of $\mathcal{D}(t)$ (see text)  
as calculated  through Eq.~(\ref{Etot}). 
Squares: under-damped regime
($\g/m = 0.27$ ps$^{-1}$ (a) and $\g/m = 4.27$ ps$^{-1}$ (b)).
Circles: over-damped regime ($\g = 27$ ps$^{-1}$ (a) $\g = 427$ ps$^{-1}$ (b)).
The dashed lines are plots of the straight lines
$\mathcal{D}(t) = 2f_{\rm eff}\g t$. 
The dotted and dot-dashed lines are stretched exponential laws
with exponent $\sigma \approx 0.22$ and $\sigma \approx 0.24$, respectively (a)
and $\sigma \approx 0.35$ and $\sigma \approx 0.48$, respectively (b).
The solid lines are plots of  formula~(\ref{BesselD}) with 
$\g_{\scs 0} = (f_{\rm eff}/d_{\rm eff}) \g$ and effective 
dimension  $d_{\rm eff} \approx  1.45$ (a), and 
$d_{\rm eff} \approx  0.83$ (b).
In the insets we plot
$\< E(t) \> - \< E(\infty) \>$ in units of $k_{\scs B}T$ (symbols).
The solid lines are the same as in the main panels.
Parameters are $T_0 = 2T$, $N=194$, $f_{\rm eff} = 0.242$, $r_c = 8$ \AA \ (a) 
and $N=309$, $f_{\rm eff} = 0.217$, $r_c = 6$ \AA \ (b).
\label{f:relax}}
\end{figure*}
%%%%%%%%%%%%%%%%%%%%%%%%%%%%%%%%%%%%%%%%%%%%%%%%%%%%%%%%

In the harmonic approximation, the total potential energy of a system of size $N$
can be written as a quadratic form 
\begin{equation}
\label{quadrform}
{U} = \frac{1}{2} (X-X^{\scs 0})^T K \, (X-X^{\scs 0})
\end{equation}
where $X=\{r_{1,x},r_{1,y},r_{1,z}, \dots, r_{N,x},r_{N,y},r_{N,z}\}$, $X^0$ 
the same vector at equilibrium and 
the ``contact'' matrix $K$ is simply  the
Hessian of the  potential energy function evaluated at the
equilibrium structure (for the sake of simplicity, we have set all masses
equal to one). Accordingly, the FP equation 
for the probability distribution  in phase space takes the form~\cite{FP}
\begin{multline}
\label{FPEq}
\frac{\de P(Y,t|Y(0))}{\de t} = \\
                  \sum_{i,j=1}^{6N}
                  \left[
                   -\mathbb{A}_{i,j} \frac{\de}{\de Y_i} Y_j
                   +\mathbb{B}_{i,j} \frac{\de^2}{\de Y_i \de Y_j}
                  \right]
                  P(Y,t|Y(0))
\end{multline}
where $Y=(X-X^{\scs 0},\dot{X})$ is the $6N$--dimensional vector of displacements
and velocities, and the matrices $\mathbb{A}$ and $\mathbb{B}$ are given by
\begin{equation}
\label{ABmatr}
\mathbb{A} = \left(
             \begin{BMAT}(e)[2pt,1cm,1cm]{c.c}{c.c} 
              0 & \mathbb{I}_{\scs 3N} \\
              -K & -\Gamma
             \end{BMAT}
             \right) \ \ \ \ \ \ 
\mathbb{B} = k_{\scs B}T \left(
             \begin{BMAT}(e)[2pt,1.2 cm,1.2 cm]{c.c}{c.c} 
              0 & 0 \\
              0 & \Gamma
             \end{BMAT}
             \right)
             \quad .
\end{equation}
The information on the coupling to the solvent of individual atoms is contained 
in the diagonal matrix $\Gamma$.
The latter has the form $\Gamma_{ij} = \gamma \delta_{ij} S_{i}$,
where the vector $S$ fixes the fraction of surface exposed to the solvent 
by each  particle ($0<S_i<1$, $i=1,2,\dots,N$). 
The parameter $\gamma$ specifies the overall strength of the viscous force.

The solution of Eq.~(\ref{FPEq}) is the multivariate Gaussian
distribution~\cite{Risken}
\begin{multline}
\label{MVGauss}
P(Y,t|Y(0)) = (2\pi)^{-3N} |\det C(t)|^{-1/2} \times \\
                     \exp \left\{
                     -\frac{1}{2}[Y-G(t)Y(0)]^TC^{-1}(t)[Y-G(t)Y(0)]
                                \right\}
\end{multline}
where $C_{i,j}(t) = \langle Y_i(t) Y_j(t) \rangle$ is the block 
correlation matrix
\begin{equation}
\label{blockcorr}
C = \left(
             \begin{BMAT}(e)[2pt,1.1cm,1.1cm]{c.c}{c.c} 
             C_{XX} & C_{X\dot{X}}  \\
             C_{X\dot{X}}  & C_{\dot{X}\dot{X}}
             \end{BMAT}
             \right)
\end{equation}
and $G(t)$ is the propagator matrix. The latter can be evaluated at any time
in terms of a normalized bi-orthogonal set of left and right eigenvectors of the 
matrix $\mathbb{A}$ (sometimes referred to as Langevin modes) as 
\begin{equation}
\label{Gbiorth}
G(t) = \Psi_R e^{\Lambda t} \Psi^T_L
\end{equation}
where $\Lambda$ is the diagonal matrix of the eigenvalues of $\mathbb{A}$,
and $\Psi_R$ and $\Psi_L$ are the matrices of right and left eigenvectors,
respectively. It is easy to show that the evolution law  for the correlations reads, 
in matrix form,
\begin{equation}
\label{cevol}
C(t) = C(\infty) + G(t)[C(0) - C(\infty)]G^T(t)
\end{equation}
where we have introduced the equilibrium correlation matrix 
\begin{equation}
\label{blockcorreq}
C(\infty) = k_{\scs B}T \left(
             \begin{BMAT}(e)[2pt,1cm,1cm]{c.c}{c.c} 
             K^{-1} & 0  \\
             0  & \mathbb{I}_{\scs 3N}
             \end{BMAT}
             \right) 
\end{equation}
$K^{-1}$ being the generalized inverse of matrix $K$.

In order  to calculate the energy relaxation to equilibrium, we first 
diagonalize matrix $\mathbb{A}$ and then evaluate the correlation
matrix~(\ref{cevol}) as a function of time 
with the aid of Eq.~(\ref{Gbiorth}).
The expression for the total energy then finally reads
\begin{equation}
\label{Etot}
%\begin{aligned}
\langle E(t) \rangle 
                     = \frac{1}{2} 
		           {\rm Tr} \, \left[
                             C_{\dot{X}\dot{X}}(t) + 
                             K C_{XX}(t) 
                           \right] \quad .                      
%\end{aligned}        
\end{equation}

To illustrate our analysis, let us examine the relaxation to equilibrium of 
a typical globular protein, Myoglobin, and of a model
Au nano-cluster with icosahedral symmetry~\cite{Siber} 
(see Fig.~\ref{f:strutt}). 
We estimate the effective fraction of surface 
exposed to the  solvent at each site (the vector $S$) 
through standard solvent-accessible surface areas.
As initial conditions, we set the specific potential energy 
at its equilibrium value $3/2k_{\scs B}T$, while each particle is given
the kinetic energy  $3/2k_{\scs B}T_{\scs 0}$ ($T_{\scs 0}>T$).
This amounts to taking in Eq.~(\ref{cevol}) 
$C_{XX}(0)=C_{XX}(\infty)=k_{\scs B}TK^{-1}$, 
$C_{X\dot{X}}(0)=C_{X\dot{X}}(\infty)=0$, and 
$C_{\dot{X}\dot{X}}(0)= k_{\scs B} T_0 \mathbb{I}_{\scs 3N}$.
Incidentally, the latter assignments, together with 
Eq.~(\ref{cevol}), also prove that  the relaxation process only depends
on the difference $\Delta T = T_{\scs 0} - T$.

Our results are summarized in Figs.~\ref{f:relax} (a) and (b),
where we plot the quantity $\mathcal{D}(t)$ defined as
\begin{equation}
\label{Dt}
\mathcal{D}(t) = -\log \left[
                        \frac{\< E(t) \> - \< E(\infty) \>}
                             {\< E(0) \> - \< E(\infty) \>}
                       \right] \quad .
\end{equation}
This representation is most useful in identifying crossovers from 
an exponential law (straight line) to a slower decay regime,
such as a stretched exponential of the form 
$\exp[-(t/\tau)^{\sigma}]$ (power law $t^\sigma$ with $\sigma<1$).

Remarkably, the two systems under scrutiny display the same 
complex behavior of the energy decay. Initially, the relaxation is exponential,
which  is the first signature of the linear damping on surface atoms.
Accordingly,  $\mathcal{D}(t) = 2f_{\rm eff}\g t/m$, where 
$f_{\rm eff} = \sum_i S_i/N$ is the total effective surface fraction of the system.
The first crossover occurs in correspondence with the onset of 
energy transfer from kinetic to potential (arrows marked C$_1$ in the plots), 
which occurs on a time scale of the order of the inverse of the maximum frequency. 
This crossover is independent of the damping rate $\g$.
At this stage the potential energies of all modes have been transferred 
a fraction of the initial kinetic energy,
and thus the relaxation process can be described in terms of Eq.~(\ref{BesselD}).
In fact, it turns out that the decay curves in the under-damped regime  
can be extremely well approximated by formula~(\ref{BesselD}) with 
$\g_{\scs 0} = (f_{\rm eff}/d_{\rm eff})\g$, where the linear surface fraction 
is naturally obtained by allowing for  an effective dimension $d_{\rm eff}$
(see insets in Figs.~\ref{f:relax} (a) and (b)). One-parameter fits
of the exact solutions yield 
$d_{\rm eff}\approx 1.45$ and $d_{\rm eff}\approx 0.83$ for Myoglobin and
the nano-cluster, respectively, which agree with the sizeable surface fraction
of the systems.

In accordance with the prescription of  formula~(\ref{BesselD}), at a time
of the order $d_{\rm eff}/f_{\rm eff}(\g/m)$ the systems undergo a second crossover to 
an {\em integrated} regime, which reflects the superposition of the time scales 
of all Langevin modes (arrows marked C$_2$ in the figures). 
This second crossover is by all means captured by the transition
from exponential to a power law implicit in formula~(\ref{BesselD}).
In the over-damped case, the fast damping of the surface atoms increases
the time scale for the dissipation of the bulk energy. As a consequence, 
the first stage of surface  relaxation stretches until the system  reaches the 
crossover to the integrated regime, thus merging the two crossovers.

In our chosen time span, the long-term behavior of the decay curves is best
approximated by a stretched exponential (see again Figs.~\ref{f:relax}). 
To this concern, it is however important to stress that asymptotically
the decay laws shall eventually bend again to a pure exponential. 
This is a finite-size effect
that reflects the progressive depletion of modes, until only the slowest one 
is left with significant energy. Consequently, its decay constant $\g_{\scs N}$ 
sets the time scale for the last exponential stage $\exp(-2\g_{\scs N}t)$, which
in both systems  is of the order of $10^{5}$ in natural
time units ($\o_0=2$). This effect may make it hard to distinguish between a
stretched exponential and a slow transition from
power-law to exponential.

In this Letter we have shown that the inhomogeneity of coupling to the solvent 
of the surface and core atoms naturally produces a hierarchy of relaxation times,
and hence complex energy relaxation,  in systems as {\em un-suspect} as harmonic 
bead-spring networks. It can be argued that non-linear effects could also affect the
nature of relaxation dynamics~\cite{straub}.
However, within our treatment, at least for small polynomial  non-linearities, an-harmonic 
effects are not expected to introduce dramatic modifications. In fact, 
the fast  damping of low-frequency modes 
with respect to  band-edge ones rapidly hinders
effective inter-mode energy exchange. This phenomenon has been well documented
in one- and two-dimensional lattices with quartic anharmonicity~\cite{Piazza1,Piazza2},
where the  crossover from exponential to a collective, power-law regime has been
observed during relaxation irrespective of the anharmonicity strength.
It is however interesting to recall that relaxation in non-linear systems is 
under certain circumstances associated with spontaneous energy 
localization~\cite{Piazza1,Aubry1}.
This phenomenon is definitely worth attention for example in the context of 
protein dynamics, as it may play a role in relaxation and redistribution 
of local energy fluctuations, such as after hydrolysis of ATP molecules.

We stress that our results are qualitatively consistent with a number 
of experimental observations~\cite{slow.clus,slow.prot}. However, 
further experimental work would be necessary in order to discriminate 
between our model and alternative explanations, such as ruggedness of 
the system energy landscape~\cite{frauenfelder}.

The authors wish to thank A. Siber for providing the coordinate file of the nano-cluster and A. Pasquarello for a critical reading 
of the manuscript.

%%%%%%%%%%%%%%%%%%%%%%%%%%%%%%%%%%%%%%%%%%%%%%%%%%%%%%%%%%%%%%%%%%%%%%%%%%%%%%%%%%

%----------------------------------------------------------------------------

\end{document}